\date{}
\documentclass[10pt,aps,prl,twocolumn,superscriptaddress]{revtex4-1}
\usepackage{graphicx,psfrag,amsmath,amssymb,amsfonts,color}
\usepackage[T1]{fontenc}
\usepackage{natbib}
\usepackage{hyperref}
\newcommand{\bra}[1]{\ensuremath{\langle{#1}|\,}}
\newcommand{\ket}[1]{\ensuremath{\,|{#1}\rangle}}

\newcommand{\xx}{\mathcal X}

\newcommand{\beq}{\begin{equation}}
\newcommand{\eeq}{\end{equation}}
\newcommand{\bea}{\begin{eqnarray}}
\newcommand{\eea}{\end{eqnarray}}

\begin{document}
\title{State dependent motional squeezing 
	of a trapped ion: new method and applications}

\author{Mart\' in Drechsler}
\affiliation{Departamento de F\'isica, FCEyN, UBA and IFIBA, UBA CONICET,
	Pabell\'on 1, Ciudad Universitaria, 1428 Buenos Aires, Argentina}
\author{M. Bel\' en Far\'ias}
\affiliation{University of Luxembourg, Physics and Materials Science Research Unit, Avenue de la Fav\'encerie 162a, L-1511, Luxembourg, Luxembourg}

\author{Nahuel Freitas}
\affiliation{University of Luxembourg, Physics and Materials Science Research Unit, Avenue de la Fav\'encerie 162a, L-1511, Luxembourg, Luxembourg}

\author{Christian T. Schmiegelow }
\affiliation{Departamento de F\'isica, FCEyN, UBA and IFIBA, UBA CONICET,
	Pabell\'on 1, Ciudad Universitaria, 1428 Buenos Aires, Argentina}
\author{Juan Pablo Paz }
\affiliation{Departamento de F\'isica, FCEyN, UBA and IFIBA, UBA CONICET,
	Pabell\'on 1, Ciudad Universitaria, 1428 Buenos Aires, Argentina}

\begin{abstract}
	We show that the 
	motion of a cold trapped ion can
	be squeezed by modulating the intensity of a phase-stable optical lattice
	placed inside the trap. As this method is reversible and
	state selective it effectively implements a 
	controlled-squeeze gate. We show how to use
	this resource, that can be useful for quantum information 
	processing with continuous variables, in order to
	prepare coherent  
	superpositions of states which are squeezed along 
	complementary quadratures. We show that these states, which
	we denote as "$\xx$-states", exhibit high sensitivity to small displacements along two 
	complementary quadratures which make them useful for quantum metrology.
\end{abstract} 

\maketitle


Cold trapped ions are one of the leading platforms for quantum
simulations~\cite{jurcevic2017direct,zhang2017observation}, quantum metrology~\cite{Ludlow2015, burd2019quantum, kienzler2016observation} and quantum information processing~\cite{kaufmann2017scalable, martinez2016real, figgatt2019parallel, tan2015multi, schafer2018fast}. 
In these devices, the preparation, manipulation and control of 
quantum states of internal and motional degrees of freedom 
plays a central role.
In this context, historical 
benchmarks have been achieved, such as the preparation and detection
of Schr\"odinger cat states~\cite{monroe1996schrodinger}, of squeezed states~\cite{meekhof1996generation} and the characterization of their decoherence~\cite{turchette2000decoherence}. 
Squeezing, an extremely
valuable resource for quantum metrology~\cite{kienzler2016observation, burd2019quantum, mccormick2019quantum} and 
information processing~\cite{gottesman2001encoding, ge2019trapped,fluhmann2019encoding}, has been 
generated in trapped ions using various methods. 
Here, we present an alternative one 
which has three main features: it is reversible,  
state selective and can generate large 
squeezings. 

Before describing our idea we review some aspects 
of the existing methods. 
In a seminal paper~\cite{meekhof1996generation}, Wineland and
coworkers demonstrated the generation of a squeezed state of motion by irradiating an ion, 
with a pair of Raman beams.  
This scheme is a variation of an older idea~\cite{cirac1993dark} that was later expanded by Home and coworkers to  generate and characterize families of 
squeezed sates~\cite{home2014science,kienzler2016observation}. These methods are based 
on the fact that squeezing is generated 
when an atom is placed in a potential modulated at  twice the trapping frequency. 
This can be achieved with ``traveling standing waves''~\cite{meekhof1996generation, kienzler2016observation} or aided by dissipation as a special kind of environmental engineering~\cite{home2014science}.
More recently Wineland and co workers implemented a method \cite{burd2019quantum} 
originally proposed in \cite{heinzen1990quantum}. In this case, the squeezing is induced by a temporal modulation of the trapping potential.  
The procedure they use is reversible. The squeezing induced by a certain modulation can always be undone by applying a second, appropriately chosen, temporal driving. Using this tool,  a small displacement was amplified by interposing it in between a squeezing and an anti-squeezing operation.

Here, we present a strategy that extends 
the above ones. 
The main idea is to place the ion in a valley (or a crest) of a optical lattice (OL) with a 
time-varying intensity. This generates a time varying 
potential that depends on the internal state of the ion, 
allowing us to squeeze the ion's motion in a state selective way.  
Implementing this method 
requires the control of the absolute phase of the lasers forming the OL. 
The ability to do this with an accuracy better than $2\%$ 
of  the lattice spacing, was demonstrated recently~\cite{schmiegelow2016phase} 
by actively stabilizing the relative phase of the OL 
by monitoring the ac-Stark shift that the same OL 
generates on the ion. 
In turn, the state dependence of the lattice potential can be obtained, by a combination of standard 
OL techniques~\cite{cohen1998nobel} and the idea of electronic shelving~\cite{nagourney1986shelved}.

\begin{figure}[b]
	\includegraphics[width=0.45\textwidth]{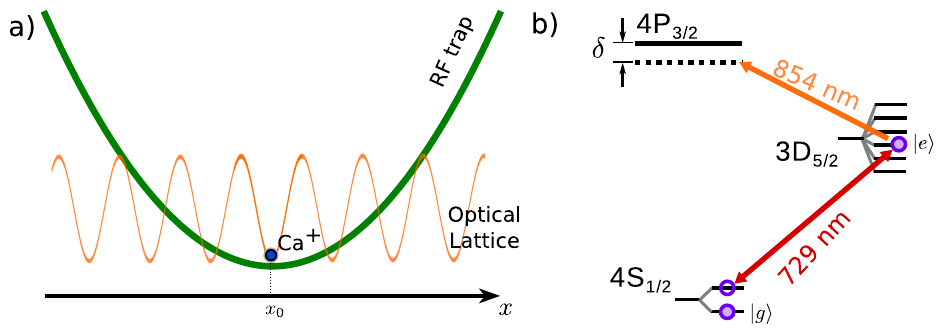}\label{fig:lattice}
	\label{fig:esquemas}
	\caption{a) Optical lattice and harmonic trapping potential, b) electronic level scheme for $^{40}$Ca$^+$. The phase of the OL is stabilized 
		so that one of its minima coincides with the one  of 
		the trapping potential. The OL has $\lambda_{OL}=
		854nm$ and is visible only when the 
		ion is in state $\ket{e}$. State dependent
		squeezing is induced by the modulation of the 
		intensity of the OL.}
\end{figure}

We show how to use this to construct a ``control-squeeze'' (C-Sqz) gate. Our method provides a new tool for quantum information processing protocols with continuous variables~\cite{gottesman2001encoding} with ion traps. 
We also show how this gate 
can be used to prepare a  
class of non-Gaussian states 
defined as coherent superpositions of 
states with squeezing along complementary quadratures. 
We denote them as ``$\xx$-states'' because 
their Wigner function is positive in an $\xx$-shaped region formed by two squeezed ellipses oriented at $90$ degrees from each 
other (these Wigner functions 
display significant oscillators in
between the positive-valued region). $\xx$-states
may be useful for quantum metrology as they are highly 
sensitive to small displacements along pairs of 
complementary quadratures (such as position and 
momentum). 

We begin by recalling that a 
squeezed state, $\ket{\xi}$, of a harmonic 
oscillator is such that $\ket{\xi}=S(\xi)\ket{0}$, where $\ket{0}$ is the ground state and $S(\xi)$ is the squeezing operator $S(\xi)=\exp\left(\left(\xi^* a^2-\xi a^{\dagger 2}\right)/2\right)$. Here, $a^\dagger$ and $a$ are creation and annihilation operators satisfying
$[a,a^\dagger]=1$ and $\xi=r e^{i\theta}$ defines the degree of squeezing $r$ and the quadrature $\theta$ along which the state is squeezed. 
The main features of a squeezing operation follow from the transformation
law:
$a'\equiv S^\dagger(\xi) a S(\xi)= 
\cosh(r) a-e^{i\theta}\sinh(r)a^\dagger$. 
For example, for $\theta=0$, we have
$(a'\pm a'^\dagger)=e^{\mp r}(a\pm a^\dagger)$. 
This shows that squeezing produces 
exponential stretching and 
exponential contraction along 
complementary quadratures.
The expansion of $\ket{\xi}$ 
in terms of energy eigenstates
$\ket{n}$ only involves even values of $n$ and reads 
\beq
\ket{\xi}={1\over{\sqrt{\cosh(r)}}}
\sum_{m\ge 0}{{\sqrt{2m!}}\over{m! 2^m}}
\left(-\tanh(r) e^{i\theta}\right)^m \ 
\ket{2m}.
\label{eq:squeezed_number_distib}
\eeq

Squeezing can be generated by varying the frequency of a harmonic oscillator..
%
%
We propose to squeeze the motion of a trapped ion by varying 
the effective 
trapping frequency with an optical lattice (OL).  
The OL can be created 
by the interference of a single laser beam 
reflected back onto itself~\cite{cohen1998nobel}.
The basic ingredients, similar to the ones used 
in~\cite{schmiegelow2016phase}, are  
shown in Figure 1. We will analyze here the case of  
$^{40}$Ca$^+$ ions, whose relevant internal states are 
shown in that Figure, but our protocol is easily portable to other ions. 
We will use an optical qubit~\cite{leibfried2003quantum} and  
choose the state $\ket{g}$ as one of the Zeeman sub-levels of 
${}^4S_{1/2}$ and $\ket{e}$ as one of the states in the 
${}^3D_{5/2}$ manifold.  
The OL will be generated with a laser detuned from the  ${}^3D_{5/2} \, \leftrightarrow \, {}^4P_{3/2}$ transition, whose wavelength is close to $\lambda_{SW} = 854nm$. We consider typical detunings of a few GHz with respect to transitions with line-widths of a few MHz. 
In such a situation the field generates an ac-Stark shift mainly for the $\ket{e}$ state, because of its coupling to the ${}^4P_{3/2}$  manifold. Conversely, the transitions from $\ket{g}$ to other levels are far 
detuned and, as a consequence, this state remains mostly unaffected by the OL. 
Thus, the Hamiltonian is the sum of the contribution 
of the harmonic  
trap, plus the one of the ac-Stark shift generated by the OL. 
It reads:
\begin{equation}
H_\text{OL}=\frac{p^2}{2m}+\frac{1}{2}m\omega_T^2
x^2+\ket{e}\bra{e} \otimes V_0\sin^2(k_lx+\Phi).
\label{eq:full-hamiltonian}
\end{equation}
Here, $\omega_T$ is the trapping frequency and 
$V_0 = \frac{\hbar\Omega_R^2}{4\delta}$ 
is the ratio between the Rabi frequency, $\Omega_R$, and the detuning 
$\delta$ ($k_l=2\pi/\lambda$ is the lattice wave vector and $\Phi$ is a phase 
that determines the position of the lattice minima with respect to the trapping potential). 
As discussed above, one can actively lock the phase to $\Phi=0$ and 
prepare a state well localized near $x=0$, where
both the OL and trapping potential have a common 
minimum.
By modulating the laser intensity (which determines 
$\Omega_R$) we introduce an explicit time dependence in the Hamiltonian (that is, $V_0$ can be transformed into $V_0(t)$).

We now analyze the evolution operators for the system 
assuming a harmonic driving of the form ${V_0(t)=\hbar\epsilon 
(1-\sin(\omega_d t-\theta))}$. 
When the ion is in state $\ket{g}$,  it only interacts with the harmonic trap and its Hamiltonian is $H_g^{(0)}=p^2/2m+m\omega_T^2x^2/2$.  
On the other hand, when the state is $\ket{e}$ the ion sees the OL. For  an initial state whose wave packet is 
concentrated near $x\approx 0$,   
the OL can be approximated by a quadratic potential
and the Hamiltonian can be written as
$H_{e}=H_e^{(0)} +H_e^{(int)}$, where $H_e^{(0)}=p^2/2m+m\omega_e^2x^2/2$ (with the rescaled trapping frequency 
$\omega_e=\sqrt{\omega_T^2
	+2k_l^2\hbar\epsilon/m}$) and 
$H_e^{(int)}=-\hbar G
\sin(\omega_dt-\theta) x^2/\sigma_e^2$, where 
the driving amplitude 
is $G=\epsilon k_l^2\sigma_e^2$ and $\sigma_e=
\sqrt{\hbar/m\omega_e}$ (this approximation
is valid in the Lamb Dicke limit where $\sigma_e\ll 
\lambda_{OL}$, and thus $\eta\equiv k_l^2\sigma_e^2\ll 1).$ 

\begin{figure}[b]
	\includegraphics[width=0.48\textwidth]{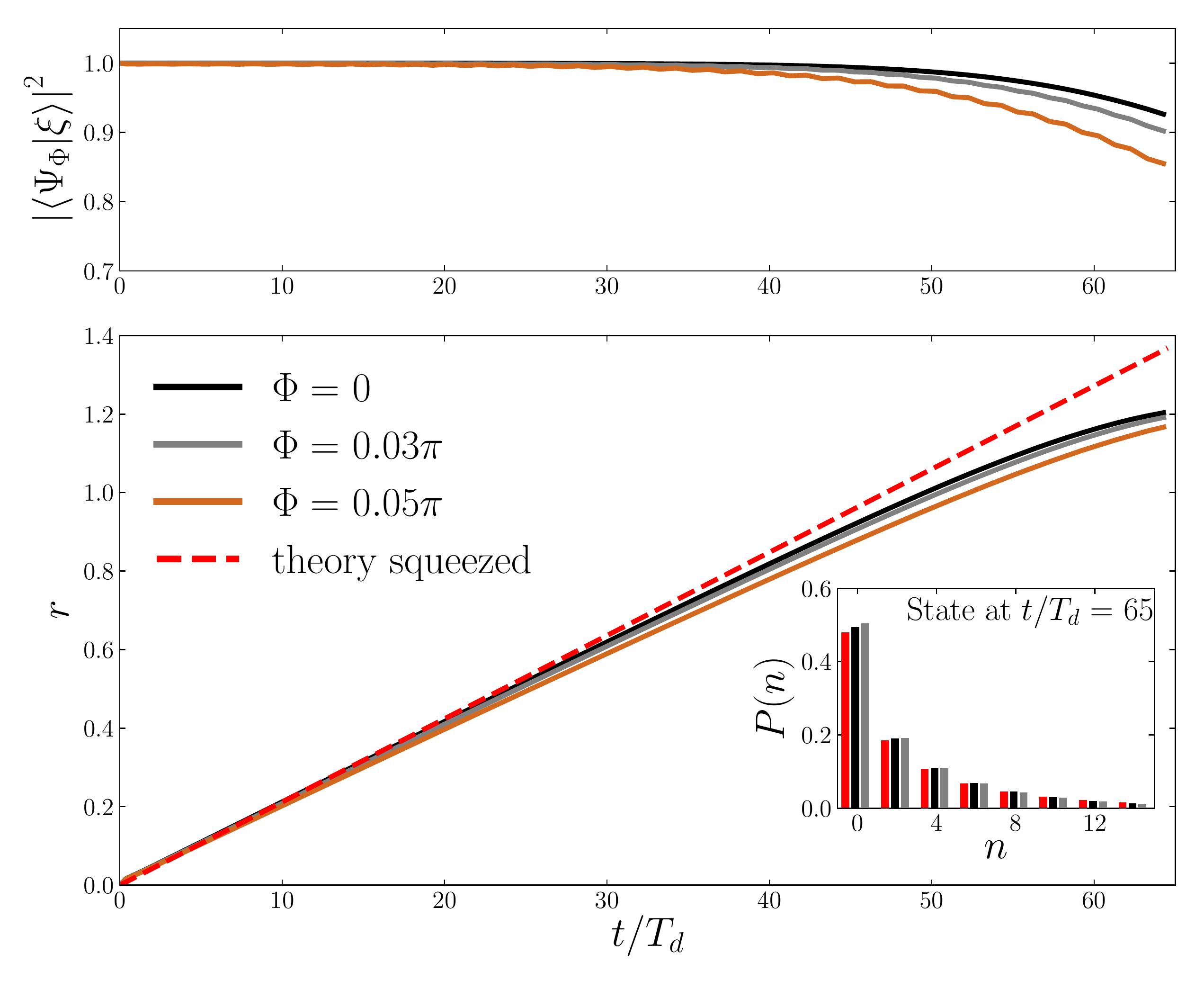}
	\caption{The squeezing of an ion
		caused by the modulation of the intensity of an optical lattice 
		(OL) of the form
		$V_0(t) =  \hbar \omega_T \left[ 1 - \sin \left( 2\omega_e  
		t) \right) \right]$. The relative position of the OL
		minimum and the one of the trap is controlled by the 
		absolute phase $\Phi$,  which is locked to $\Phi=0$ up to 
		experimental errors whose effect are seen in the Figure. 
		The obtained state coincides with the one
		predicted by theory, which is
		$\ket{\xi}$ with $\xi=Gt/2$. Deviations are seen, as
		expected, for long times (when the nonlinearity of the 
		OL potential becomes significant) or for large  
		values of $\Phi$. Errors induce the decay 
		the overlap with the ideal state and deviations
		from the predicted phononic population $P(n)$.
		\label{fig:squeeze_evolution}
	}
\end{figure}

In parametric resonance, 
when $\omega_d=2\omega_e$, 
squeezing is generated 
in a constructive way. To prove this, we write the Hamiltonian $H_e$ in the interaction picture with respect to $H_e^{(0)}$ 
(and in the rotating wave approximation, RWA) 
as $\tilde H_e=i{{\hbar G}\over 4}(a_e^2e^{-i\theta}-a_e^{\dagger 2}e^{i\theta})$. Here $a_{e}$ is 
the canonical operator 
associated with $H_{e}^{(0)}$, which can be interchanged
with $a_g$ (the one of $H_g^{(0)}$) in the 
Lamb Dicke limit (since $a_e\approx 
a_g+{\mathcal O}(\eta)$).
Then, the evolution operator for a time $T$ is 
$\tilde U(T)=S({{GT}\over 2}
e^{i\theta})$, 
%
which would generate a squeezed state with a degree 
of squeezing that grows linearly with time. 
Clearly, this squeezing 
can be inverted by an identical modulation 
with a temporal phase $\pi-\theta$. 

To benchmark our approximation we numerically solved the 
Schr\"odinger equation considering the full non-linearity of the OL potential 
and avoiding the RWA invoked above, i.e. the Hamiltonian of Eq. \eqref{eq:full-hamiltonian}. 
The results, shown in Figure~\ref{fig:squeeze_evolution}, agree with the above analysis: the overlap between the 
numerical
state and the squeezed state with $\xi=Gt/2$
stays close to unity even for small variations of the phase $\Phi$ consistent with experimentally achievable errors 
(which can be as small as $2\%$).  
After nearly $100$ driving periods $r$
becomes of order unity for a modulation 
amplitude  
$\epsilon
\approx \omega_T$  ($\epsilon>\omega_T$ was
attained in~\cite{vonlindenfels2019spin} for 
an OL on an S-P transition in Calcium).


Using the above results we can write the following expression
for the full evolution operator (in the 
interaction picture associated with $H_g^{(0)}$):
%
%
\beq
U(T)=\ket{g}\bra{g}\otimes I+
\ket{e}\bra{e} \otimes  U_{ge}\ 
S(\xi_T),
\eeq
where $\xi_T={{GT}\over 2}e^{i(\theta)}$ (notice
that the operator $U_{ge}=U_g^\dagger(T) U_e(T)$  
maps the interaction 
picture associated with $H_e^{(0)}$ into the
one of $H_g^{(0)}$).
This shows that by modulating the OL we implement a  
controlled-squeezing operation, which
we will denote as C-Sqz$(r,\theta)$: 
the motional state does not change
if the internal state is $\ket{g}$ while it is 
squeezed by $S(\xi)$ when the internal state is $\ket{e}$
(where $\xi=re^{i\theta}$). 

Let us now show how to use C-Sqz$(r,\theta)$   
to prepare a special class of non-Gaussian states.
When the internal state is either $\ket{e}$ or $\ket{g}$, this operator is Gaussian.
However, when combined with rotations of the internal state it can be used to prepare highly non-Gaussian states, which 
are an essential requirement to achieve universality in
quantum computation with continuous variables. 
We will prepare even and odd $\xx$--states, which are
defined as the following superposition
of squeezed states: 
$\ket{\xx_\pm}=N_\pm(\ket{\xi}\pm\ket{\xi e^{i\pi}})$ with 
the normalization constant $N_\pm=1/\sqrt{2\pm2/\sqrt{\cosh(2r)}}$.

These states can be prepared using the following six-step protocol (we consider $\theta=0$, i.e. $\xi=r$): 
i) Prepare the ion in the motional ground state and in a balanced superposition of the internal
state: ${1\over{\sqrt{2}}}(\ket{g}+\ket{e})\otimes\ket{0}$ ; 
ii) Apply a C-Sqz(r,0) and generate the state 
${1\over{\sqrt{2}}}(\ket{g}\otimes\ket{0}+\ket{e}\otimes 
U_{ge}\ket{r})$ ; 
iii) Perform a $\pi$ rotation in the internal state to obtain
${1\over{\sqrt{2}}} (\ket{e}\otimes\ket{0}+\ket{g}\otimes U_{ge}\ket{r})$;
iv) Apply C-Sqz$(r,\pi)$ to obtain 
${1\over{\sqrt{2}}}(\ket{e}\otimes U_{ge}\ket{-r}+\ket{g}\otimes U_{ge}
\ket{r})$; 
v) Perform a $\pi/2$ rotation in the internal state to 
obtain  
${1\over 2}({1\over{N_+}} \ket{e}\otimes U_{ge}
\ket{\xx_+}+{1\over {N_-}}\ket{g}\otimes U_{ge} \ket{\xx_-})$;
vi) Measure the internal state and obtain either $\ket{e}$ 
or $\ket{g}$. These two results appear respectively with
probabilities $1/4N_\pm^2$. In each case, 
the motional state (in the interaction picture of $H_e^{(0)}$) 
is either
$\ket{\xx_+}$ or $\ket{\xx_-}$.
Note that for large $r=GT/2$ both 
states are equally likely but for smaller values of $r$ the
even state (which is a superposition of $n=0,4,8,..$
states) is much more likely to be prepared than the 
odd state (that only contains $n=2,6,10,...$).

\begin{figure}[t!]
	\includegraphics[width=0.48\textwidth]{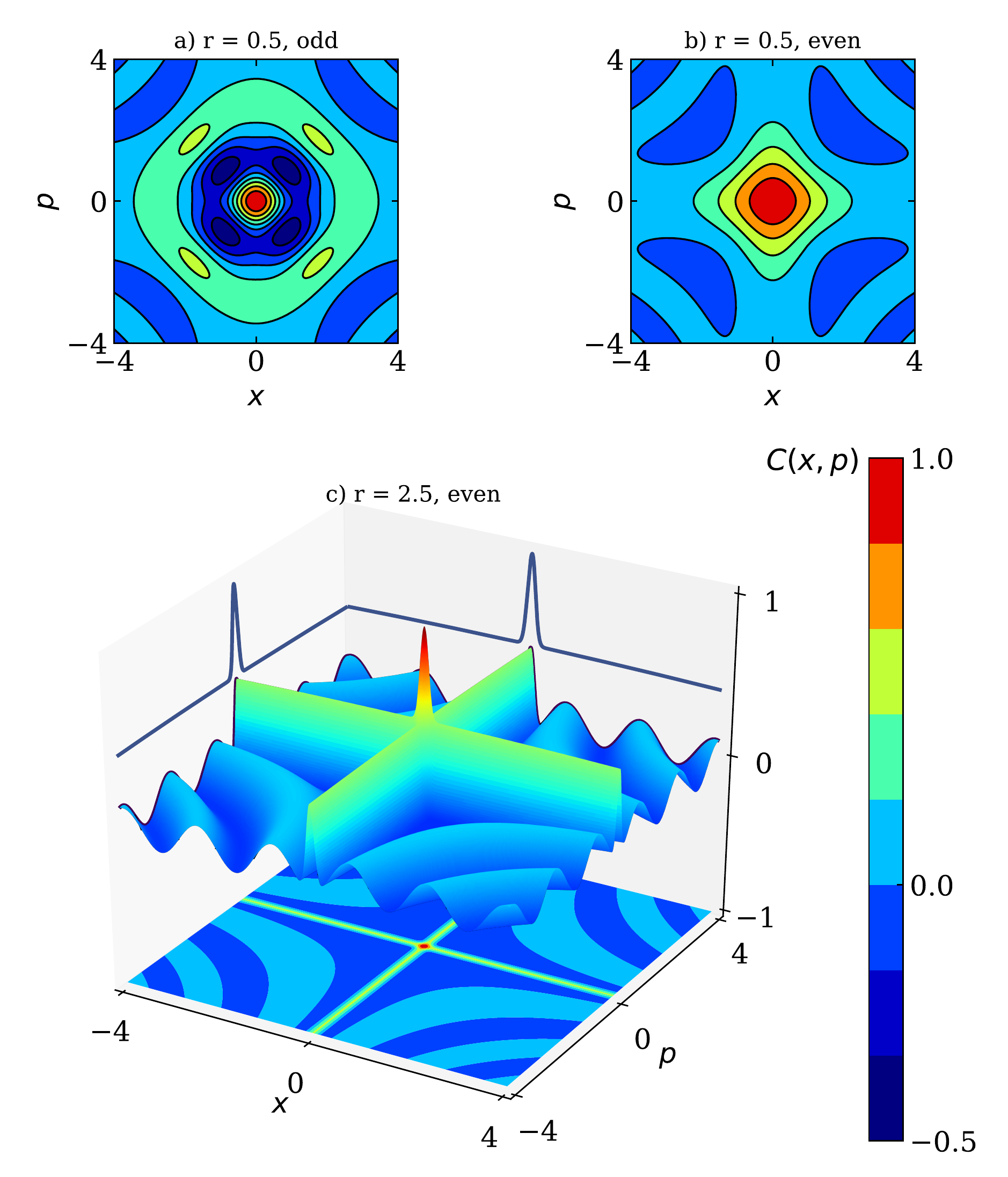}
	\caption{Characteristic function for $\xx$--states. a) and b) show the results for odd and even $\xx$-states for low squeezing ($r=0.5$) where the $\xx$--shape is barely visible . For the odd state $\xx_-$ oscillations close to the origin with negative values are seen.  These make this state orthogonal to 
		the ground state. For higher squeezing, as in c) where $r=2.5$, the $\xx$--shaped region is clearly visible. Also one sees $C(x,p)$ rapidly decays close to the origin where it 
		becomes zero for displacements along the 
		diagonals with magnitude that scales as 
		$e^{-r}$.
		\label{fig:characteristc_functions}
		}
\end{figure}

The even and odd $\xx$-states have very interesting metrological properties.  To visualize them, we analyze their Wigner function $W(\alpha)$. In fact, as the states have well defined
parity, $W(\alpha)$ is proportional to the simpler characteristic function 
$C_\pm(\alpha)=\bra{\xx_\pm}D(\alpha)\ket{\xx_\pm}$, where 
$D(\alpha)=\exp(\alpha a^\dagger -\alpha^* a)$ is
a displacement operator (thus, one can show
that $W(\alpha)\propto C(2\alpha)$).
%
Using, $\alpha=x+ip$, we found $C(\alpha)$ can be written as:
\begin{widetext}
	\begin{eqnarray}
	C_\pm(x,p)=2N_\pm^2 
	\left(e^{-{{(x^2+p^2)\cosh(2r)}\over 2}}
	\cosh\left({{x^2-p^2}\over 2}\sinh(2r)\right)
	\pm{{e^{-{{(x^2+p^2)}\over{2\cosh(2r)}}}}
		\over{\sqrt{\cosh(2r)}}} 
	\cos\left(xp\tanh(2r)\right)\right),
	\label{eq:characteristic_function}
	\end{eqnarray}
\end{widetext}

The first term in the r.h.s. is the sum of the two
direct terms while the second one brings about the 
interference and the oscillations. 
In Figure~\ref{fig:characteristc_functions} we display $C(\alpha)$ for $\xx$--states
with small and large squeezing. For large $r$ the 
result is simple to interpret: 
the $\xx$--region becomes exponentially large 
and narrow extending along the two
main quadratures (with a high peak in the origin, where  
$C_\pm(0)=1$). The oscillations are oriented along hyperbolae located in the four quadrants. 

The behavior of $\ket{\xx_-}$ is remarkable: as this 
state is orthogonal to the ground state. $C_-(\alpha)$ decays very fast for small $|\alpha|$ 
and becomes negative inside the unit circle (that defines the Gaussian
support of the ground state). Using Equation \eqref{eq:characteristic_function} we find its zeros. These indicate
the displacements that are required to transform 
$\ket{\xx_-}$ into an orthogonal state.  
Evaluating $C_-(x,p)$ on the diagonal lines 
defined by the equation 
$x^2=p^2$, we find that 
${C_-(x,p)=0}$ iff 
$
\sqrt{\cosh(2r)}e^{-{{x^2\sinh^2(2r)}\over{\cosh(2r)}}}
=
\cos(x^2\tanh(2r))$. For large $r$ 
the solution to this equation is close
to $x^2\approx re^{-2r}$. 
This shows the extreme sensitivity of 
$\ket{\xx_-}$ to small displacements along both
main diagonals. To the contrary, for the $\xx_+$-state, $C_+$
only vanishes for $x\approx 1$. 
Similarly, the behavior along the two main quadratures (where either $x=0$ or $p=0$) is such that when $x\approx e^{-r}$, $C(\alpha)$ rapidly decays to half its maximum value and stays constant (until large values of $x$ are reached). This is different to the behavior of an ordinary squeezed state where  there is a fast decay along one axis ($x$) and a slow decay along the other one. 

Finally, we note that one could also conceive $\xx$--states with 
more than a single ion. With two ions, for example, they
are superpositions of two-mode states which are 
squeezed along complementary quadratures. To generate
them we need a C-Sqz for two modes, which can be 
implemented extending the previous idea. For example, 
we can trap two ions in a linear Paul trap and orient the OL so that it 
affects the motion along one of the transverse directions illuminating 
both ions at once (the two ions are placed at common 
minima of the trapping and OL potentials). Then by beating the laser intensity with two frequencies that excite the parametric resonances of
both normal modes we would generate state dependent squeezing
of the two modes at once. In 
particular, a two-mode squeezed state would be created when the 
two beating  signals are dephased by $\pi$. This  is a  
generalized C-Sqz gate for two modes that can be used to  
build a simple sequence of operations that would 
create a generalized $\xx$--state (we omit this sequence and simply mention that such states are superpositions of two EPR states with complementary properties).

We presented a method to create state dependent squeezing
and explored one application: the generation of displacement-sensitive non-Gaussian $\xx$-states.   Such tools are critical in the development of new techniques which will allow the use of the motional modes of trapped ions for metrological applications as well as for realizing quantum information protocols with continuous variables. 

We acknowledge supports of grants from ANPCyT, UBACyT and CONICET (Argentina) as well as discussions with Ferdinand Schmidt-Kaler and David Wineland.

 
%

\end{document}